# Implementation of the Tangent Sphere and Cutting Plane Methods in the Quantitative Determination of Ligand Binding Site Burial Depths in Proteins Using FORTRAN 77/90 Language

**Vicente M. Reyes, Ph.D.***

**E-mail:** vmrsbi.RIT.biology@gmail.com

*work done at:

Dept. of Pharmacology, School of Medicine,
University of California, San Diego
9500 Gilman Drive, La Jolla, CA 92093-0636

&

Dept. of Biological Sciences, School of Life Sciences
Rochester Institute of Technology
One Lomb Memorial Drive, Rochester, NY 14623

**Abbreviations:** CP, cutting plane; TS, tangent sphere; CPM, cutting plane method; TSM, tangent sphere method; CPi, CP or CPM index; TSi, TS or TSM index; BS, binding site; LBS, ligand BS; GC, global or molecular centroid; LC, local centroid; PDB, Protein Data Bank

## 1. ABSTRACT:

The degree of ligand burial depth gives an indication of a receptor protein's flexibility, as the extent of receptor conformational change required to bind a ligand is thought to vary directly with depth of burial in the bound state. In a companion paper (Reyes, V.M. 2015a), we report on the Tangent Sphere (TS) and Cutting Plane (CP) methods - two complementary methods to quantify, in a size-independent manner, the degree of ligand binding site (LBS) burial in a protein receptor - and their applications. In this report, we present results that demonstrate the effectiveness of the several FORTRAN 77 and 90 source codes used in the implementation of the two related procedures, as well as the precise implementation of the procedures themselves. Particularly, we show here that application of the TS and CP methods on a theoretical model protein in the form of a spherical grid of points accurately portrays the expected behavior of the TS and CP indices, the predictive parameters obtained from the two methods, respectively. We additionally show that results of the implementation of the TS and CP methods on six real protein receptors from the dataset of Laskowski et al. (1996) are in general agreement with their findings regarding cavity sizes in these proteins. The six FORTRAN program source codes we present in this report are: (1.) find_molec_centr.f, (2.) tangent_sphere.f, (3.) find_CP_coeffs.f, (4.) CPM_Neg_Side.f, (5.) CPM_Pos_Side.f and (6.) CPM_Zero_Side.f. The first program calculates the x-, y- and z-coordinates of the molecular geometric centroid of the protein, a.k.a., global centroid (GC), which will be the center of the TS. Its radius is the distance between the GC and the local centroid (LC), which is the centroid of the bound ligand or portion of the LBS. The second program finds the number of protein atoms inside, outside and on the TS itself. The third program determines the four coefficients A, B, C and D of the equation of the CP, $Ax + By + Cz + D = 0$. The CP is the plane tangent to the TS at GC (and thus normal to the line segment connecting GC and LC). The fourth, fifth and sixth programs determine the number of protein atoms lying on the negative side, positive side, and on the CP itself. The process is illustrated in Figures 1A and 1B (*ibid.,* 2015a.).

## 2. INTRODUCTION:

Most proteins carry out their biological functions in the cell either by binding to small organic molecules called ligands, or by binding to other proteins, in processes generally termed protein-ligand interactions (PLI) and protein-protein interactions (PPI), respectively. The two procedures discussed in this paper, namely the Tangent Sphere Method (TSM) and the Cutting Plane Method (CPM; Reyes, V.M., 2015a; Cheguri, S. & Reyes, V.M., 2011), were inspired by the initial works by Ben-Shimon, A., et al., (2005); they both deal with PLI (Reyes, V.M., 2015a, b & c; Reyes, V.M. & Sheth, V.N., 2011), but may be applied to PPI as well (Reyes, V.M., 2015d). TSM and CPM are complementary, but not redundant, methods of quantifying the degree of burial of a ligand or LBS in the receptor protein. The initial information needed are the protein centroid coordinates, which we call the global centroid (GC), and the atomic coordinates of the bound ligand or those of the amino acids comprising the ligand binding site (LBS), from which the local centroid (LC) may be calculated. The equations of the CP and the TS are then derived from these coordinates. The quantification is independent of the size of the protein, therefore the degree of ligand burial in receptor proteins of differing sizes may be directly compared. The degree of ligand or LBS burial is related to the TS index (TSi) and CP index (CPi) determined by the TSM and CPM, respectively. The TSi is the percentage of protein atoms inside the TS, while the CPi is the percentage of protein atoms on the external side of the CP (the side opposite the global centroid). Knowledge of the degree of burial of a ligand in its receptor protein is important, as it gives one an idea about the extent of protein conformational change that the protein undergoes upon binding of the ligand. The deeper the ligand or its BS, the greater the protein conformational change required to bind it, and conversely, the shallower the ligand or its BS, the lesser the proteinconformational change required to bind it.

The TSM and CPM have also been used as an auxiliary confirmatory test in the 3D tetrahedral search motif method for predicting specific ligand binding sites in proteins (Reyes, V.M., 2015b & 2015c), as well as the 3D interface search motif tetrahedral pair method for the prediction of protein-protein interaction partners (Reyes, V.M., 2015d), using the double-centroid reduced representation of proteins (Reyes, V.M. & Sheth, V.N., 2011).

## 3. DATASETS AND METHODS:

The main dataset upon which the programs presented in this paper can be applied is the Protein Data Bank (PDB). The PDB is the main international repository for protein 3D structures solved experimentally either by x-ray crystallography or protein NMR (Berman et al., 2000). To demonstrate the usefulness and efficacy of the procedures described here, the programs were originally applied to the dataset in Laskowski (1996), composed of 67 globular (roughly spherical) monomeric enzymes with bound ligand(s). The procedure was also applied to a "theoretical" protein made by constructing a 3D grid of points in the shape of a sphere of radius 50 units and with center at the origin, with the points themselves representing protein atoms (Reyes, V.M., 2015a). We recommend that the present paper be read in conjunction with aforementioned paper in order for the reader to see precisely how the programs presented here are applied.

The minimum requirements in running the Fortran program source codes reported here is a UNIX computing environment and a Fortran 77/90 compiler software as all Fortran program source codes must be compiled before they are run. Source program codes presented in this work were written in either Fortran 77 (Holoien, M.O. & Behforooz, A., 1991; Mayo, W. & Cwiakala, M., 1994; and Nyhoff, L. & Leestma, S., 1996) or Fortran 90 ( Nyhoff, L. & Leestma, S., 1996 & 1999; Metcalf, M. & Reid, J.K., 1999; and Chapman, S.J., 1997). In order to apply the procedure in high-throughput batch mode, UNIX C-shell (Powers, S. et al., 2002; Anderson, G. & Anderson, P., 1986; and Birns, P. et al., 1985) as well as Perl (Tisdall, J., 2001; and Berman, J.J., 2007) scripts were written. In some complex cases, the scripts were constructed using text manipulation by sed & awk (Dougherty, D. & Robbins, A., 1997; and Aho et al., 1988).

## 4. RESULTS AND DISCUSSION:

Tabulated below are the Fortran program source codes presented in this paper. The page numbers refer to the pages in the present paper where the program starts. We refer the reader to our previous paper, Reyes, V.M., 2015a, for the implementation and application of these programs. Specifically, please refer to Figures 1A and 1B of said paper.

**Table of Programs**

---------------------------------------------------------------------------------------------------------

-------------------------------------------------------------

Most protein PDB files have water molecule coordinates, and these are initially removed by appropriate UNIX commands. Program 1, find_molec_centr.f, finds the molecular (geometric) centroid of the protein by averaging the x-, y- and z-coordinates of all atoms in the PDB file of the protein. The result is a single point, ($x_c$, $y_c$, $z_c$), representing the geometric centroid of the protein molecule. We term it 'geometric' centroid because no weighting was done to account for the different atomic weights of the different protein atoms. The geometric centroid is the global centroid (GC). The same program may be used to calculate the local centroid (LC), which is the centroid of the ligand or LBS; it is assumed that some or all of the coordinates of the ligand or LBS are known, which can then be inputted into Program 1. Program 2, tangent_sphere.f, effectively constructs the TS: it takes in the GC coordinate as center and the distance between the GC and LC coordinates as radius of the TS. It then outputs the number of protein atoms inside, outside and on the TS. Unsurprisingly, it is quite rare that atoms are found right on the TS itself. Program 3, find_CP_coeffs.f, calculates the coefficients A, B, C and D of the equation of the CP, $Ax + By + Cz + D = 0$, from the GC and LC coordinates and based on the definition of the CP (see Reyes, V.M., 2015a). Program 4, CPM_Neg_Side.f, determines the number of protein atoms on the negative side ($Ax + By + Cz + D < 0$) of the CP, while Program 5, CPM_Pos_Side.f, finds the number of protein atoms on the positive side $Ax + By + Cz + D > 0$). Finally, program 6, CPM_Zero_Side.f, calculates the number of protein atoms lying on the CP itself. Again, unsurprisingly, it is rare that protein atoms are found to lie on the CP itself, so this number is almost always zero. Figures 2A, 2B and 2C (*ibid.*) show results from the case of the theoretical spherical protein, while the rest of the Figures (*ibid.*) show results from the dataset in Laskowski et al. (1996).

### 4.1 Implementation of the Programs on a Theoretical Model Protein.

The concept behind the TS and CP methods are illustrated in Figure 1 using the spherical model protein (a.k.a. theoretical protein). The center of the TS in this case coincides with that of the model protein (grid of points). The ligand or ligand BS is assumed to be at the bottom of the pit representing the ligand binding pocket and where it touches the TS. The CP is shown as the blue line tangent to the TS at the same point. As one goes from the left protein to the right, the LBS depth increases; at the same time, the TS becomes smaller and smaller, and the CP comes nearer and nearer the center of the protein. The TS index (TSi) is defined as the percentage of protein atoms lying inside the TS. The CP index (CPi) is defined as the percentage of protein atoms lying on the external side (the side of the plane opposite the one containing the global centroid (GC). Note that as the LBS goes from shallow to medium to deep (spherical model protein from left to right), the TSi decreases and the CPi increases. But that is only half of the story. When the LBS is deep enough that the GC actually lies on the CP, the behavior of the TSi and the CPi actually reverses as the LBS continues to get deeper.

The four curves in Figure 2 shows the behavior of TSi and CPi (in two different scenarios: 2x2 = 4 curves) versus distance of LBS along the z-axis for the spherical model protein. In the first scenario, the blue and red curves show the behavior of TSi and CPi, respectively, as the global centroid (GC) goes incrementally from the LBS opening, the 'north pole' at (0,0,50), to the opposite pole, the 'south pole' at (0,0,-50). As expected the TSi (blue curve) starts at 100%, decreases, then reaches 0% at the global centroid (GC, the centroid of the sphere, at which the TS is just a point), then increases again, and reaches 100% when the LC reaches the opposite pole. The CPi (red curve) on the other hand starts at 0%, increases, then reaches a maximum of 50% when the LC coincides with the GC at the center of the spherical protein. In the second scenario, we assume that the LBS opening is unknown, as this is usually the case with real protein structures (or known but not considered, because it does not lend itself to automation). The green and maroon curves corresponding to the TSi and CPi, respectively, show how they behave in the second scenario. This essentially is simply the left half of the blue+red curve but with twice the number of data points. This is due to the fact that the right half of the blue+red curve had simply "folded into" its left half due to the ambiguity in the direction of where the LBS opening is located relative to the LC.

### 4.2 Implementation of the Programs on Real Proteins.

We then tried to implement our procedure to real proteins, which, of course, are not perfectly spherical as the model protein we described above. Table 1 shows the identities of the six proteins, A-F, from the 67 proteins in the Laskowski et al. dataset (1996) that were tested in this small pilot study. Protein A, (PDB ID: 1TDE), thioredoxin reductase from *E. coli*, has a single bound ligand, FAD-500. Protein B, (PDB ID: 1RPA), rat acid phosphatase, has three ligands, namely, TAR-343, NAG-344 and NAG-347. Protein C, (PDB ID: 1XNB), xylanase from *B. circulans/T. harzianum*, also has a single ligand, S04-191. Protein D, (PDB ID: 1HNE), human neutrophil elastase, has five ligands, which are ALM I-5, MSU I-1, ALA I-2, ALA I-3 and PRO I-4. Protein E, (PDB ID: 1PDA), phorphobilinogen from E. coli, has two ligands, DPM-314 and ACY-315. Finally, protein F, (PDB ID: 4BCL a.k.a. 3EOJ), FMO protein from *P. aestuarii* has seven ligands, namely, BCL-1, BCL-2, BCL-3, BCL-4, BCL-5, BCL-6 and BCL-7.

Table 2 shows the results upon implementation of the TS and CP methods on the above proteins as to their respective LBS burial depths. Note that three submethods, namely ligand ("lig"), residue ("res") and sidechain ("sdc"), were employed in each TSi and CPi determination. The difference between these three submethods are described in detail in the accompanying paper (Reyes, V.M. 2015a), suffice it to say that they are all used for the calculation of the LC coordinates. For rigor and reliability, values are reported only if all three submethods agree in their LBS burial depth prediction as to "shallow," "medium," or "deep." Also, note that the LBS depths were evaluated for each of the ligands in the six proteins, i.e., if a protein has n bound ligands, the LBS depths of each of the n ligands were calculated. In this small dataset, n = 1, 3, 1, 5, 2 and 7 for proteins A, B, C, D, E and F, respectively.

From the table, we can see that protein A's single ligand, FAD, is predicted to be deep, with an average (over the three submethods, lig, res, sdc) TSi of around 3.0, and an average CPi of about 36.0. Protein B has three ligands, one TAR and two NAG molecules. Only for the TAR and one of the NAG molecules did the three submthods agree: ligand TAR was predicted deep, with an average TSi of 5.0 and CPi of 26.0; one NAG ligand was predicted to be shallow, with an average TSi of 89.0 and CPi of 1.5; no index is reported for the second NAG ligand where the three submethods did not unanimously agree. Meanwhile, protein C has a single ligand, $SO_4$, and is predicted as being shallow, with an average TSi of 82.0 and CPi of 0.6. Protein D has four bound ligands, but only one of them, MSU, was predicted to be shallow by all three submethods, with an average TSi of 85.0 and CPi of 1.0. The other four, ALM, two ALA molecules and PRO, were indeterminate as the three submethods did not agree. Protein E has two ligands, DPM and ACY, and both are predicted to be deep, with an average TSi's of 6.0 and 3.4, and CPi's of 22.0 and 28.0, respectively. Finally, for protein F, which has seven bound molecules of BCL, only three of the ligands are predicted with consistency by the three submethods of our procedure, namely two deep with an average TSi's of 1.0 and 2.5, and CPi's of 32.0 and 28.0, respectively; and one medium, with an average TSi of 20.0 and CPi of 14.5. No indices are reported for the other four BCLs as the three submethods are not unanimous in their assessments.

### 4.3 Conclusion

In conclusion, we conclude that the six FORTRAN program source codes presented in this paper do perform their expected functions successfully. We also conclude that both the TS and CP methods are effective and robust. Our objective for the foreseeable future is to apply the procedures described in this and the companion paper (Reyes, V. M., 2015a) in large-scale, especially to all receptor proteins known to bind specific ligands and whose structures have been experimentally solved or computationally predicted.

## 5. ACKNOWLEDGMENT:

This work was supported by an Institutional Research and Academic Career Development Award to the author, NIGMS/NIH grant number GM 68524. The author wishes to acknowledge the San Diego Supercomputer Center, the UCSD Academic Computing Services, and the UCSD Biomedical Library, for the help and support of their staff and personnel. He also acknowledges the Division of Research Computing at RIT, and computing resources from the Dept. of Biological Sciences, College of Science, at RIT.

## 8. FIGURES and LEGENDS:

---

---

## 9. TABLES and LEGENDS:

---

---

## 9. PROGRAMS:

**Program 1:**

**########## Start of Pogram "find_molec_centr.f" ##########**

```
program  find_molecular_centroid


! c c c c c c c c c c c c c c c c c c c c c c c c c c c c c c c c c c c c c
!                                                                         c
```

```
!   Author:  Vicente M. Reyes, Ph.D.                                  c
!            Dept. of Pharmacol., Skaggs Sch. of Pharm. & Pharm. Sci.   c
!         &  Dept. of Integrative Biosci., S.D. Supercomptr. Ctr.       c
!            La Jolla, CA  92093-0505  U.S.A.                            c
!                                                                        c
! c c c c c c c c c c c c c c c c c c c c c c c c c c c c c c c c c c c c c

      character*30 left
      character*30 right
      integer count
      real x, y, z, sum_x, sum_y, sum_z, xc, yc, zc

      open (unit =1, file = "filei")
      open (unit =2, file = "fileo")

      count = 0
      sum_x = 0.00
      sum_y = 0.00
      sum_z = 0.00

888   read(1,100,end=333) left, x, y, z, right
100   format(A30, f8.3, f8.3, f8.3, A30)

      count = count + 1
      sum_x = sum_x + x
      sum_y = sum_y + y
      sum_z = sum_z + z

      go to 888

333   continue

      xc  = sum_x/count
      yc  = sum_y/count
      zc  = sum_z/count

      write (2,200) xc, yc, zc
200   format(f10.5, f10.5, f10.5)

!     print*, "x = ", xc
!     print*, "y = ", yc
!     print*, "z = ", zc

      close(2)
      close(1)

      stop
      end
```

**##########   End of Pogram  “find_molec_centr.f”  ##########**

**Program 2:**

**##########   Start of Pogram   “tangent_sphere.f”   ##########**

```
c  program: tangent_sphere.f


c c c c c c c c c c c c c c c c c c c c c c c c c c c c c c c c c c c c c c c
c                                                                             c
c   Author:  Vicente M. Reyes, Ph.D.                                          c
c            Dept. of Pharmacol., Skaggs Sch. of Pharm. & Pharm. Sci.        c
c          & Dept. of Integrative Biosci., S.D. Supercomptr. Ctr.            c
c            La Jolla, CA  92093-0505  U.S.A.                                 c
c                                                                             c
c c c c c c c c c c c c c c c c c c c c c c c c c c c c c c c c c c c c c c c

      character*30 left
      character*22 mesg
      character*5 lbl1,lbl2,lbl3,lbl4,lbl5,lbl6,lbl7,lbl8
      real count, inside, outside, pct_in, pct_out
      real xc, yc, zc , xp, yp, zp, radius, dist

!     input files:
      open (unit =1, file = "filea")  ! <-- center of sphere: x,y,z:  ??.prot.CM
      open (unit =2, file = "fileb")  ! <-- radius of sphere: ??.lbs?.prot.CM.len
      open (unit =3, file = "filec")  ! <-- pdb file of protein under test:
??.prot

!     output file:
      open (unit =4, file = "filew")  ! <-- atoms inside sphere
      open (unit =5, file = "filex")  ! <-- atoms outside sphere
      open (unit =6, file = "filey")  ! <-- atoms right on sphere
      open (unit =7, file = "filez")  ! <-- results: %in, %out, %on

      count = 0.0
      inside = 0.0
      outside = 0.0
      righton = 0.0

      lbl1 = 'dist='
      lbl2 = '%in= '
      lbl3 = '#in= '
      lbl4 = '%out='
      lbl5 = '#out='
      lbl6 = '%on= '
      lbl7 = '#on= '
      lbl8 = 'Tot= '

      mesg = "point right on sphere!"

888   read(1,100) xc, yc, zc
100   format (f10.5, f10.5, f10.5)

      read(2,200) radius
200   format (f10.5)

999   read(3,300,end=444) left, xp, yp, zp
300   format(A30, f8.3, f8.3, f8.3)

      count = count + 1.0

      dist = sqrt((xc-xp)**2 + (yc-yp)**2 + (zc-zp)**2)

      if(dist.lt.radius) then
```

```
            write(4,400) left, xp, yp, zp, lbl1, dist
            inside = inside + 1.0

      elseif (dist.gt.radius) then
            write(5,400) left, xp, yp, zp, lbl1, dist
            outside = outside + 1.0

      else
            write(6,401) left, xp, yp, zp, lbl1, dist, mesg
            righton = righton + 1.0


400   format(A30, f8.3, f8.3, f8.3, 3x, A5, f10.5)
401   format(A30, f8.3, f8.3, f8.3, 3x, A5, f10.5, 3x, A16)

      endif

      go to 999

444   pct_in = (inside/count)*100.0

      pct_out = (outside/count)*100.0

      pct_on = (righton/count)*100.0

      write(7,700) lbl2,pct_in, lbl3,inside, lbl4,pct_out,
     +  lbl5,outside, lbl6,pct_on, lbl7,righton, lbl8,count

700   format (A5, f9.3, 3x, A5,f9.0, 3x, A5,f9.3, 3x, A5,
     +    f9.0, 3x, A5,f5.3, 3x, A5,f4.0, 3x, A5,f9.0)
brodie:/home/vmrsbi/projects/TSM.theoret (378) % c TSM.f
c  program: tangent_sphere.f

      character*30 left
      character*22 mesg
      character*5 lbl1,lbl2,lbl3,lbl4,lbl5,lbl6,lbl7,lbl8
      real count, inside, outside, pct_in, pct_out
      real xc, yc, zc , xp, yp, zp, radius, dist

!     input files:
      open (unit =1, file = "filea")  ! <-- center of sphere: x,y,z:  ??.prot.CM
      open (unit =2, file = "fileb")  ! <-- radius of sphere: ??.lbs?.prot.CM.len
      open (unit =3, file = "filec")  ! <-- pdb file of protein under test:
??.prot

!     output file:
      open (unit =4, file = "filew")  ! <-- atoms inside sphere
      open (unit =5, file = "filex")  ! <-- atoms outside sphere
      open (unit =6, file = "filey")  ! <-- atoms right on sphere
      open (unit =7, file = "filez")  ! <-- results: %in, %out, %on

      count = 0.0
      inside = 0.0
      outside = 0.0
      righton = 0.0

      lbl1 = 'dist='
      lbl2 = '%in= '
      lbl3 = '#in= '
      lbl4 = '%out='
      lbl5 = '#out='
```

```
      lbl6 = '%on= '
      lbl7 = '#on= '
      lbl8 = 'Tot= '

      mesg = "point right on sphere!"

888   read(1,100) xc, yc, zc
100   format (f10.5, f10.5, f10.5)

      read(2,200) radius
200   format (f10.5)

999   read(3,300,end=444) left, xp, yp, zp
300   format(A30, f8.3, f8.3, f8.3)

      count = count + 1.0

      dist = sqrt((xc-xp)**2 + (yc-yp)**2 + (zc-zp)**2)

      if(dist.lt.radius) then

            write(4,400) left, xp, yp, zp, lbl1, dist
            inside = inside + 1.0

      elseif (dist.gt.radius) then
            write(5,400) left, xp, yp, zp, lbl1, dist
            outside = outside + 1.0

      else
            write(6,401) left, xp, yp, zp, lbl1, dist, mesg
            righton = righton + 1.0


400   format(A30, f8.3, f8.3, f8.3, 3x, A5, f10.5)
401   format(A30, f8.3, f8.3, f8.3, 3x, A5, f10.5, 3x, A16)

      endif

      go to 999

444   pct_in = (inside/count)*100.0

      pct_out = (outside/count)*100.0

      pct_on = (righton/count)*100.0

      write(7,700) lbl2,pct_in, lbl3,inside, lbl4,pct_out,
     +  lbl5,outside, lbl6,pct_on, lbl7,righton, lbl8,count

700   format (A5, f9.3, 3x, A5,f9.0, 3x, A5,f9.3, 3x, A5,
     +    f9.0, 3x, A5,f5.3, 3x, A5,f4.0, 3x, A5,f9.0)

      close(7)
      close(6)
      close(5)
      close(4)
      close(3)
      close(2)
      close(1)

      stop
```

```
      end
```

########## End of Pogram "tangent_sphere.f" ##########

**Program 3:**

########## Start of Pogram "find_CP_coeffs.f" ##########

```
c  program find_CP_coeffs.f

c c c c c c c c c c c c c c c c c c c c c c c c c c c c c c c c c c c c c c
c                                                                         c
c   Author:  Vicente M. Reyes, Ph.D.                                      c
c            Dept. of Pharmacol., Skaggs Sch. of Pharm. & Pharm. Sci.     c
c         &  Dept. of Integrative Biosci., S.D. Supercomptr. Ctr.         c
c            La Jolla, CA  92093-0505  U.S.A.                             c
c                                                                         c
c c c c c c c c c c c c c c c c c c c c c c c c c c c c c c c c c c c c c c

c  see ~/science_notes, lines 2515 ff

      character*45 n_mesg, p_mesg, z_mesg
      real p,q,r,s,t,u
      real A,B,C,D,val

      open (unit =1, file = "filea")  !this is point #1, lbs?.CM
      open (unit =2, file = "fileb")  !this is point #2, prot.CM

      open (unit =3, file = "fileo")  !output file: prot CM on pos. side of CP

      p_mesg = 'gCM on (+) side of CP => external side is (-)'
      n_mesg = 'gCM on (-) side of CP => coeff signs reversed'
      z_mesg = 'gCM lies right on the CP!! A very rare case!!'

888   read(1,100) p, q, r   ! this is P1: lbs.CM  == local CM
999   read(2,100) s, t, u   ! this is P2: prot.CM  == global CM

100   format(f10.5, f10.5, f10.5)


!*******************************************************************
! complex in question; i.e.,   L = (p,q,r) = local CM    and
!                              G = (s,t,u) = global CM
!                              Q = (x,y,z) = any point on the cutting plane, CP
!
!
! The equation of the plane is then:    Ax + By + Cz + D = 0
!
! where  A = s-p
!        B = t-q
!        C = u-r
! and    D = p(p-s) + q(q-t) + r(r-u)                                QED
!
! NOTE:  The prot CM (global CM) is always on the positive side of the CP and
```

```
!        thus the exteternal side of the CP is alweays its negative side!!!!!!!
!
!*********************************************************************

      A = s-p
      B = t-q
      C = u-r
      D = ((p*(p-s)) + (q*(q-t)) + (r*(r-u)))

      val = A*s + B*t + C*u + D       !!!  it can be shown that "val" is always (+)

      if (val.gt.(0.0)) then

      write(3,300) A,B,C,D,p_mesg, val
300   format(f12.5, f12.5, f12.5, f12.5, 3x, A45, f10.3)

      elseif (val.lt.(0.0)) then

      write(3,300) -A,-B,-C,-D,n_mesg, -val

      elseif (val.eq.(0.0)) then

      write(3,300) A,B,C,D,z_mesg, val

      endif

333   continue

      close(3)
      close(2)
      close(1)

      stop
      end
```

**##########   End of Pogram  “find_CP_coeffs.f”  ##########**

**Program 4:**

**##########   Start of Pogram   “CPM_Neg_Side.f “  ##########**

```
c  program: CPM_NegSid.f

c c c c c c c c c c c c c c c c c c c c c c c c c c c c c c c c c c c c c c
c                                                                        c
c   Author:  Vicente M. Reyes, Ph.D.                                     c
c            Dept. of Pharmacol., Skaggs Sch. of Pharm. & Pharm. Sci.    c
c         &  Dept. of Integrative Biosci., S.D. Supercomptr. Ctr.        c
c            La Jolla, CA  92093-0505  U.S.A.                            c
c                                                                        c
c c c c c c c c c c c c c c c c c c c c c c c c c c c c c c c c c c c c c c


      character*30 left
      character*5 lbl1, lbl4, lbl5, lbl8
      real numneg, pct_neg, val
```

```
      real A, B, C, D, x, y, z, total

!     input files:
      open (unit =1, file = "filea")  ! <-- CP coeffs: ??.lbs?.CP_coeffs
      open (unit =2, file = "fileb")   ! <-- pdb file of protein under test:
??.prot

!     output file:
      open (unit =4, file = "filey")  ! <-- prot atoms on (-) side of CP
      open (unit =5, file = "filez")  ! <-- results: %(-), #(-)

      total = 0.0
      numneg = 0.0

      lbl1 = 'val= '
      lbl4 = '%(-)='
      lbl5 = '#(-)='
      lbl8 = 'Tot= '

888   read(1,100) A,B,C,D
100   format(f12.5, f12.5, f12.5, f12.5)

999   read(2,300,end=444) left, x, y, z
300   format(A30, f8.3, f8.3, f8.3)

      total = total + 1.0

      val = A*x + B*y + C*z + D

      if (val.lt.(0.0)) then

            write(4,400) left, x, y, z, lbl1, val
            numneg = numneg + 1.0

400   format(A30, f8.3, f8.3, f8.3, 3x, A5, f10.5)

      endif

      go to 999

444   pct_neg = (numneg/total)*100.0

      write(5,500)lbl4,pct_neg, lbl5,numneg, lbl8,total

500   format(A5,f10.4, 3x, A5,f10.0, 3x, A5,f10.0)

      close(5)
      close(4)
      close(2)
      close(1)

      stop
      end
```

**#########    End of Pogram   "CPM_Neg_Side.f" #########**

**Program 5:**

**##########  Start of Pogram "CPM_Pos_Side.f" ##########**

```
c  program: CPM_PosSid.f

c c c c c c c c c c c c c c c c c c c c c c c c c c c c c c c c c c c c c c
c                                                                         c
c   Author:  Vicente M. Reyes, Ph.D.                                      c
c            Dept. of Pharmacol., Skaggs Sch. of Pharm. & Pharm. Sci.     c
c         &  Dept. of Integrative Biosci., S.D. Supercomptr. Ctr.         c
c            La Jolla, CA  92093-0505  U.S.A.                             c
c                                                                         c
c c c c c c c c c c c c c c c c c c c c c c c c c c c c c c c c c c c c c c

      character*30 left
      character*5 lbl1, lbl2, lbl3, lbl8
      real numpos, pct_pos, val
      real A, B, C, D, x, y, z, total

!     input files:
      open (unit =1, file = "filea")  ! <-- CP coeffs: ??.lbs?.CP_coeffs
      open  (unit  =2,  file  =  "fileb")   !  <--  pdb  file  of  protein  under  test:
??.prot

!     output file:
      open (unit =3, file = "filex")  ! <-- prot atoms on (+) side of CP
      open (unit =5, file = "filez")  ! <-- results: %(+), #(+)

      total = 0.0
      numpos = 0.0

      lbl1 = 'val= '
      lbl2 = '%(+)='
      lbl3 = '#(+)='
      lbl8 = 'Tot= '

888   read(1,100) A,B,C,D
100   format(f12.5, f12.5, f12.5, f12.5)

999   read(2,300,end=444) left, x, y, z
300   format(A30, f8.3, f8.3, f8.3)

      total = total + 1.0

      val = A*x + B*y + C*z + D

      if (val.gt.(0.0)) then

            write(3,400) left, x, y, z, lbl1, val
            numpos = numpos + 1.0

400   format(A30, f8.3, f8.3, f8.3, 3x, A5, f10.5)

      endif

      go to 999
```

```
444   pct_pos = (numpos/total)*100.0

      write(5,500)lbl2,pct_pos, lbl3,numpos, lbl8,total

500   format(A5,f10.4, 3x, A5,f10.0, 3x, A5,f10.0)

      close(5)
      close(3)
      close(2)
      close(1)

      stop
      end
```

########## End of Pogram "CPM_Pos_Side.f" ##########

**Program 6:**

########## Start of Pogram "CPM_Zero_Side.f" ##########

```
c  program: CPM_ZerSid.f

c c c c c c c c c c c c c c c c c c c c c c c c c c c c c c c c c c c c c c
c                                                                         c
c   Author:  Vicente M. Reyes, Ph.D.                                      c
c            Dept. of Pharmacol., Skaggs Sch. of Pharm. & Pharm. Sci.     c
c         &  Dept. of Integrative Biosci., S.D. Supercomptr. Ctr.         c
c            La Jolla, CA  92093-0505  U.S.A.                             c
c                                                                         c
c c c c c c c c c c c c c c c c c c c c c c c c c c c c c c c c c c c c c c

      character*30 left
      character*5 lbl1, lbl6, lbl7, lbl8
      real numzer, pct_zer, val
      real A, B, C, D, x, y, z, total

!     input files:
      open (unit =1, file = "filea")  ! <-- CP coeffs: ??.lbs?.CP_coeffs
      open  (unit  =2,  file  =  "fileb")   !  <--  pdb  file  of  protein  under  test:
??.prot

!     output file:
      open (unit =3, file = "filex")  ! <-- prot atoms on (0) side of CP
      open (unit =5, file = "filez")  ! <-- results: %(0), #(0),

      total = 0.0
      numzer = 0.0

      lbl1 = 'val= '
      lbl6 = '%(0)='
      lbl7 = '#(0)='
      lbl8 = 'Tot= '

888   read(1,100) A,B,C,D
100   format(f12.5, f12.5, f12.5, f12.5)

999   read(2,300,end=444) left, x, y, z
```

```
300   format(A30, f8.3, f8.3, f8.3)

      total = total + 1.0

      val = A*x + B*y + C*z + D

      if (val.eq.(0.0)) then

            write(3,400) left, x, y, z, lbl1, val
            numzer = numzer + 1.0

400   format(A30, f8.3, f8.3, f8.3, 3x, A5, f10.5)

      endif

      go to 999

444   pct_zer = (numzer/total)*100.0

      write(5,500)lbl6,numzer, lbl7,pct_zer, lbl8,total

500   format(A5,f10.4, 3x, A5,f10.0, 3x, A5,f10.0)

      close(5)
      close(3)
      close(2)
      close(1)

      stop
      end
```

**########## End of Pogram “CPM_Zero_Side.f” ##########**

**10. FIGURES:**

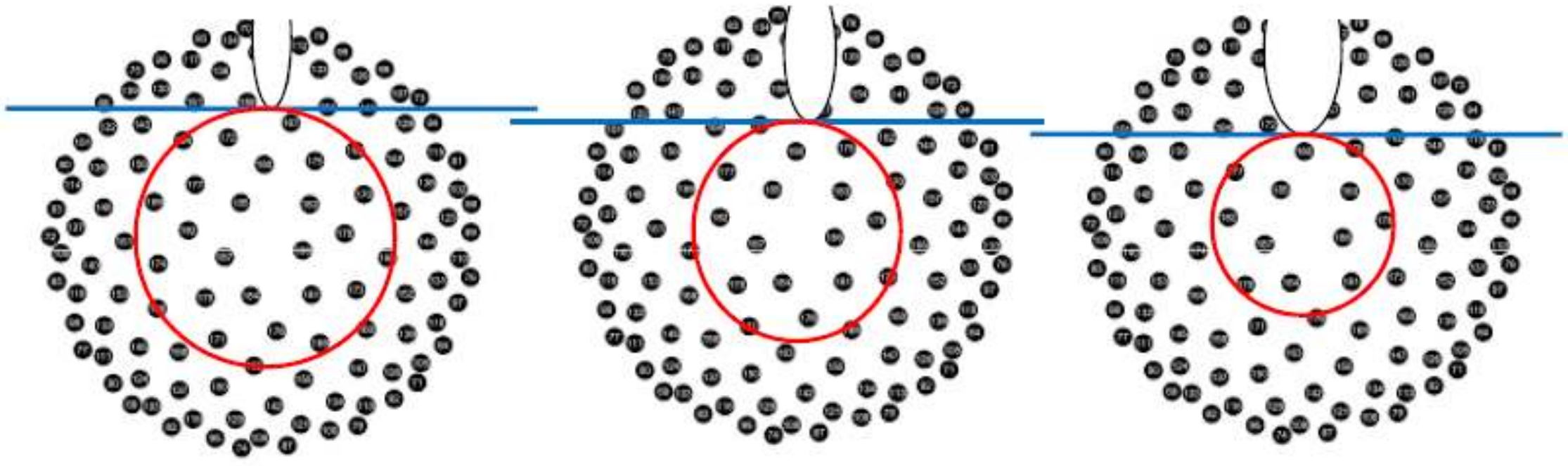

**Figure 1.**

Figure 1 illustrates the concept behind the Tangent Sphere and Cutting Plane methods. A hypothetical protein is represented by a collection of black dots collectively forming a sphere; the tangent sphere (TS) is represented by the red circle, whose center is the protein's centroid; the cutting plane (CP) is represented by the blue line, which is tangent to the TS at the ligand binding site (LBS). As the LBS gets deeper (from left to right), the volume of the TS decreases, while the portion of the protein cut off by the CP increases. The TSA index (TSi) is the number of protein atoms inside the TS, while the CP index (CPi) is the number of protein atoms cut off by the CP.

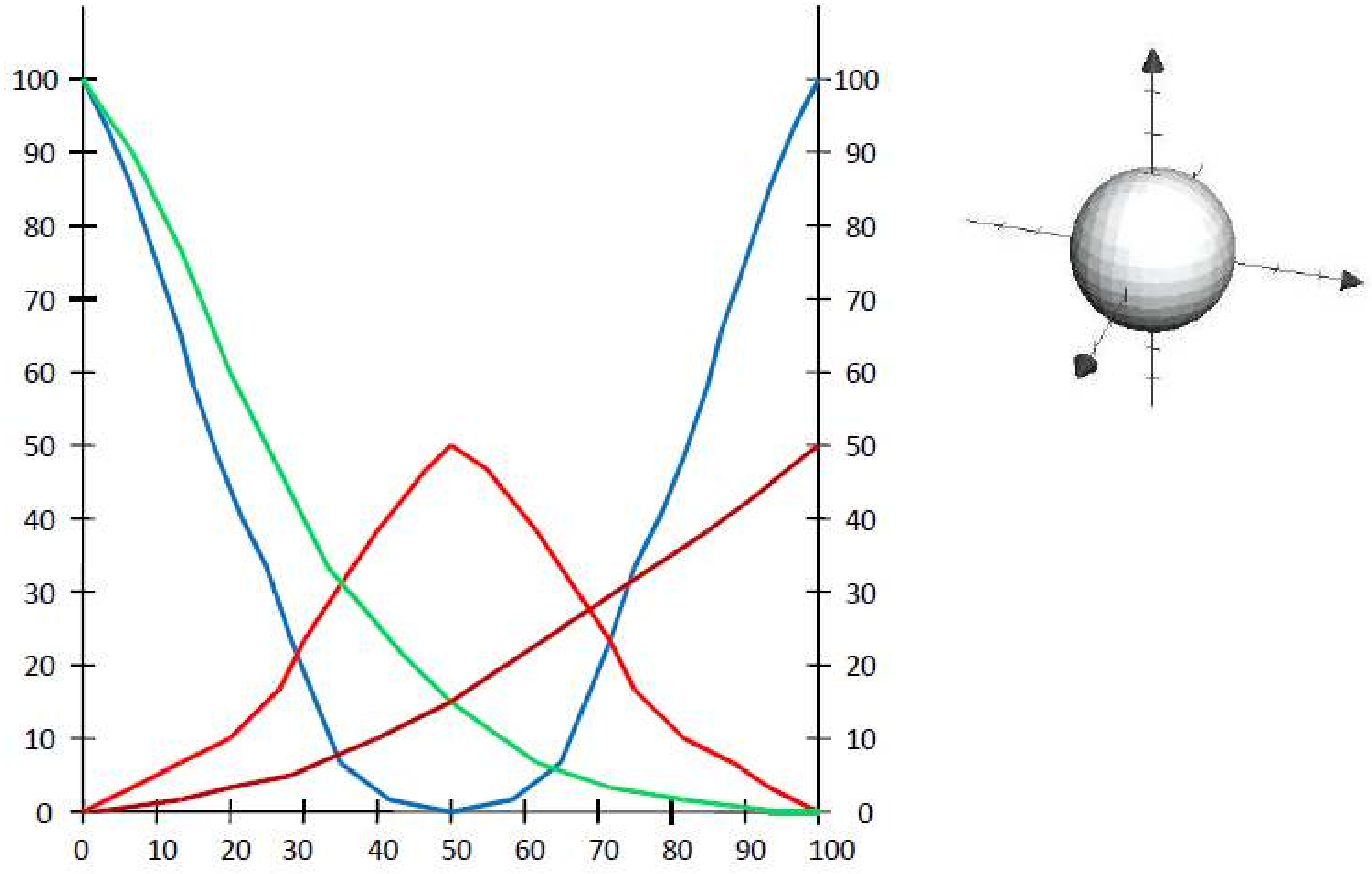

**Figure 2.**

On The plots on the left panel in Figure 2 refer to the theoretical (a.k.a. model) protein shown schematically on the right. The model protein is a three-dimensional grid of points that collectively form a sphere with center at (0,0,0,) and radius 50 units, and separated by equal distances of 1.5 units along the x-, y- and z-directions. The LBS opening is designated to be at the "north pole" of the sphere, i.e., at (0,0,50). The horizontal axis of the plot in the left panel represent the distance from the LBS along the z-axis, and the vertical axis represent percentages of protein atoms that are either inside the TS (blue curve) or cut off (intercepted, or on the external side of) by the CP (red curve). If the LBS opening (from which the ligand 'enters' the protein LBS) is not considered (as in real cases, since this information is usually unknown, or unamenable to batch-mode computation) only the left half of the blue+red curves will be displayed, but with double the number of data points (because the right half of the blue+red curves would have "folded into" the left half. These are the green curve and maroon curves, for the TS and CP methods, respectively.

**11. TABLES:**

**Table 1.**

| Protein | Abbrev.[1] | PDB ID | # ligands | Description |
| --- | --- | --- | --- | --- |
| A | Ae | 1TDE | 1 | Thioredoxin Reductase from *E. coli* |
| B | Au | 1RPA | 3 | Rat Acid Phosphatase |
| C | Bc | 1XNB | 1 | Xylanases from *B. Circulans* and *T. Harzianum* |
| D | Bm | 1HNE | 5 | Human Neutrophil Elastase |
| E | Cg | 1PDA | 2 | Porphobilinogen Deaminase from *E. coli* |
| F | Co | 4BCL[2] | 7 | Fenna–Matthews–Olson protein from *P. aestuarii* |

[1] in reference paper, Reyes, V.M., 2015x
[2] same as protein 3EOJ (FMO protein from *P. aestuarii)*

Table 2 identifies the six test proteins we used in this mini-study to test the efficacy of the TS and CP methods in quantifying the depth of burial of the ligand or ligand binding site in a receptor protein. The abbreviations in the second column refer to those in our main paper regarding this procedure (Reyes, V.M., 2015a), where these six proteins were part of the 67 test proteins taken from the dataset of Laskowski et al., 1996. Their PDB IDs are shown in the third column; note that the number of bound ligands vary among the proteins; please refer to the PDB website (www.rcsb.org) for more details about these structures.

| Protein | Abbrev. (lig #) | TS Index lig | TS Index res | TS Index sdc | CP Index lig | CP Index res | CP Index sdc | Conclusion re. LBS |
|---|---|---|---|---|---|---|---|---|
| A | Ae(1) | 1.02 | 3.88 | 5.48 | 39.86 | 34.15 | 35.20 | deep |
| B | Au(1) | 6.26 | 5.22 | 3.51 | 24.80 | 26.59 | 27.89 | deep |
| | Au(2) | 96.99 | 82.75 | 90.44 | 0.39 | 2.83 | 1.75 | shallow |
| | Au(3) | n/a | n/a | n/a | n/a | n/a | n/a | n/a |
| C | Bc(1) | 97.93 | 82.16 | 74.93 | 0/00 | 0.48 | 0.83 | shallow |
| D | Bm(1) | 90.89 | 77.08 | 84.47 | 0.37 | 2.08 | 0.98 | shallow |
| | Bm(2) | n/a | n/a | n/a | n/a | n/a | n/a | n/a |
| | Bm(3) | n/a | n/a | n/a | n/a | n/a | n/a | n/a |
| | Bm(4) | n/a | n/a | n/a | n/a | n/a | n/a | n/a |
| | Bm(5) | n/a | n/a | n/a | n/a | n/a | n/a | n/a |
| E | Cg(1) | 3.61 | 7.44 | 4.45 | 24.05 | 21.92 | 21.47 | deep |
| | Cg(2) | 3.38 | 2.90 | 4.05 | 26.16 | 30.07 | 26.50 | deep |
| F | Co(1) | 0.00 | 2.09 | 0.44 | 37.10 | 23.68 | 32.43 | deep |
| | Co(2) | 0.66 | 4.55 | 4.52 | 33.49 | 27.39 | 26.07 | deep |
| | Co(3) | 22.83 | 21.27 | 15.73 | 16.29 | 14.96 | 18.45 | medium |
| | Co(4) | n/a | n/a | n/a | n/a | n/a | n/a | n/a |
| | Co(5) | n/a | n/a | n/a | n/a | n/a | n/a | n/a |
| | Co(6) | n/a | n/a | n/a | n/a | n/a | n/a | n/a |
| | Co(7) | n/a | n/a | n/a | n/a | n/a | n/a | n/a |

**Table 2.**

Table 2 shows the results of the application of the TSM and CPM on the six test proteins picked out at random from the 67 proteins in the dataset of Laskowski et al., 1996. For those structures with more than one bound ligand, the TSM and CPM had to be performed for each bound ligand, e.g., three times for protein B, five times for protein D, twice for protein E and seven times for protein F. The three TS indices using three methods of implementing the TSM (lig = ligand method; res = residue method; sdc = sidechain method) are shown in columns three, four and five. The three CP indices using three methods of implementing the CPM (lig, res, sdc) are shown in columns six, seven and eight. The conclusions based on the TS and CP indices are shown in column nine.